\documentclass{elsart}
\date{20 October 2004}

\usepackage[latin1]{inputenc}
\usepackage{natbib}
\usepackage{color}
\definecolor{grey}{rgb}{.85,.85,.85}
\usepackage{graphicx}
\usepackage{amssymb}

\newcounter{revcomm}

\graphicspath{{./fig/}}

\newcommand{\li}[1]{\mbox{#1}}
\newcommand{\equnit}[1]{\mbox{#1}}
\newcommand{\up}[1]{\textsuperscript{#1}}
\newcommand{\pb}[0]{\pagebreak{}}

\newcommand{\NumUnitDist}[0]{~\hspace{-.3ex}}
\newcommand{\eqdist}[0]{~\hspace{-.4ex}}
\newcommand{\prc}[0]{~\hspace{-.5ex}\%}
\newcommand{\uwlf}{\ensuremath{
    {\NumUnitDist\mbox{W(m\eqdist{}K)\textsuperscript{-1}}}}}
\newcommand{\udens}{\ensuremath{
    {\NumUnitDist\mbox{kg\eqdist m\up{-3}}}}}
\newcommand{\umeter}[0]{\ensuremath{
    {\NumUnitDist\mbox{m}}}}
\newcommand{\upres}[0]{\ensuremath{
    {\NumUnitDist\mbox{MPa}}}}
\newcommand{\uvel}[0]{\ensuremath{
    {\NumUnitDist\mbox{m\eqdist s\textsuperscript{-1}}}}}
\newcommand{\uinch}[0]{\ensuremath{
    {\mbox{\lq\lq}}}}

\begin{document}

\begin{frontmatter}
 
  \title{Thermal Conductivity from Core and Well log Data}
  \author{Andreas Hartmann\corauthref{lab1}}
  \ead{Andreas@geophysik.rwth-aachen.de}
  \corauth[lab1]{Corresponding author}
  \author{\hspace{-1ex}, Volker Rath, }
  \author{Christoph Clauser}
  
  \address{Department for Applied Geophysics, RWTH Aachen University,
    Lochnerstr.  4-20, 52056 Aachen, Germany}
%\ead[url]{http://www.geophysik.rwth-aachen.de}

\begin{abstract}   
  The relationships between thermal conductivity and other
  petrophysical properties have been analysed for a borehole drilled
  in a Tertiary Flysch sequence. We establish equations that permit
  us to predict rock thermal conductivity from logging data. A regression
  analysis of thermal conductivity, bulk density, and sonic velocity
  yields thermal conductivity with an average accuracy of better than
  0.2\uwlf{}.  As a second step, logging data is used to compute a
  lithological depth profile, which in turn is used to
  calculate a thermal conductivity profile. From a comparison of the
  conductivity-depth profile and the laboratory data it can be
  concluded that thermal conductivity can be computed with an accuracy
  of less than 0.3\uwlf{} from conventional wireline data.  The
  comparison of two different models shows that this approach can be
  practical even if old and incomplete logging data is used. The
  results can be used to infer thermal conductivity for boreholes
  without appropriate core data that are drilled in a similar
  geological setting.
   
\end{abstract}

\begin{keyword}
petrophysics \sep thermal conductivity \sep well logging
\end{keyword}

\end{frontmatter}
%\tableofcontents{}

% main text
\section{Introduction}
\label{intro}
    The German Molasse Basin is currently analysed with
geothermal methods in order to determine groundwater flow rates in the
deep subsurface \cite{Clauser2002}. This requires to determine
accurately and separate from each other the effects of the different
heat transport processes: Steady-state conductive, transient
conductive, and advective heat flow. Computing the steady-state
conductive temperature field requires detailed thermal conductivity
data. Two requirements should be met: First, lateral variations in
thermal conductivity have to be known accurately for regional studies.
Second, a detailed analysis of a particular borehole requires a high
resolution thermal conductivity profile. Core material in that area is
either confined to a few boreholes or available only for a particular
layer. This violates both requirements. Logging data, on the other
hand, offer better spatial coverage and depth resolution. In a
situation with an incomplete coring sequence this helps to prevent
biased sampling in lithologies not representative of the full
sequence. It is therefore highly desirable to derive thermal
properties from logging data to improve the sparse geothermal
database.

Methods to compute thermal conductivity from wireline logs can be
classified into two main categories \cite{Beardsmore2001}. The first
approach relates one or more logging measurements or some derived
property directly to thermal properties via empirical relationships.
This method has been used to compute thermal conductivity in several
studies \cite{Doveton1997,Vacquier1988,Evans1977,
  Goss1975_prediction} and a review is given in \cite{Blackwell1989}.
In the second approach, the major mineral or rock components are
identified and the volumetric fractions of these components are
derived from regular wireline data. This composition together with
component thermal conductivity values is then used to compute the
effective thermal conductivity assuming an appropriate mixing law
\cite{Williams1990,Brigaud1990,Demongodin1991}.  This approach is more
flexible than the first one.  For instance, it does not require the
same suite of logs in each borehole as long as they are suitable to
compute component volumes.

It has been pointed out \cite{Goss1975_prediction,Blackwell1989} that
results of the current methods are confined to a geographic region or
geological setting. In the following we would like to assess the two
different methods discussed above for the Molasse Basin. As the area
was explored for hydrocarbons in the past, logging data often are old
and comprise only sonic or natural gamma logs. Although it would be
more desirable to use only the second approach, it is necessary to
evaluate direct methods as well to make full use of the available
logging data.  We test direct methods on a set of core data, that was
analysed in the laboratory and the compositional approach using the
same set of laboratory data together with wireline data obtained in
the borehole.  
%\alt{Mineral thermal conductivities for rock
%forming minerals generally vary from about 2\uwlf{} (for example
%feldspars and clay minerals) to 7.7\uwlf{} (crystalline
%quartz)~\cite{Clauser1995a}. Water on the other hand as the most
%common pore fluid has value of 0.6\uwlf{}. Thus porosity is the most
%important factor controlling bulk thermal conductivity.  \alt{This is
%  particularly true high porosity and uniform lithology.}  For
%instance, in marine, high-porosity sediments thermal conductivities
%might be computed from a porosity log only \cite{Villinger1994}. For
%rocks with porosities of less than 20 to 30\prc{}, the influence of
%porosity decreases and rock composition needs to be taken into account
%\cite{Griffiths1992}.}

%\section{Core and well log data}
\section{Core data}
\label{sec:lab_meas}

\begin{figure}[htbp]
  \centering
  \includegraphics[width=14cm,trim=0 10 0 0,clip]
  {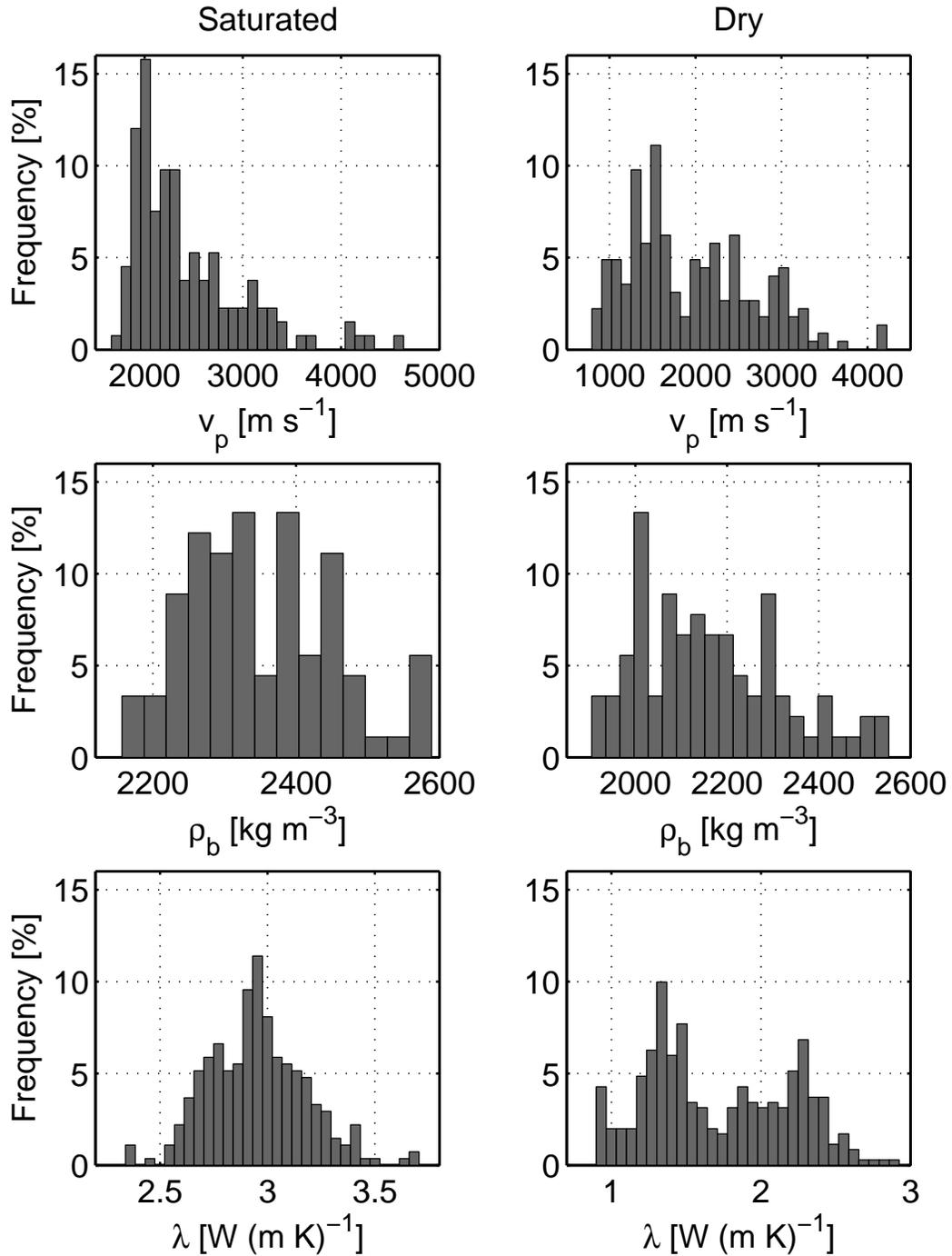}
  \caption{Histograms of sonic velocity $v_p$, bulk density $\rho_b$,
    and thermal conductivity $\lambda$ measured in the laboratory on
    core samples. Properties measured are from top to bottom: sonic
    velocity, bulk density, and thermal conductivity. Values are shown
    for saturated (left) and dry (right) conditions. The bimodal
    distribution corresponds to the occurrence of two different
    lithologies in the borehole (see also
    figure~\ref{fig:bw_composite_qc}).}
  \label{fig:bw_lab_hist}
\end{figure}

The data set for laboratory analysis is obtained from a borehole
drilled for hot water into the Upper Marine Molasse formation. This
formation consists of shaly sandstones and marls. The borehole was
cored between 570--810\umeter{} depth with nearly 100\prc{} recovery.
The sequence consists of a succession of shaly sandstones and marl
beds.  Porosity ranges from 10--30\prc{}.  Split cores were available
for measurements in sections of 1\umeter{} length.  In general, we
analysed every fifth section, but in some depth intervals we studied
all cores. The cores had broken into several pieces during coring and
splitting. We selected two to five samples from each core section for
further analysis. The samples were dried at 60°C to prevent cracking
or alteration of clay minerals.  We measured thermal conductivity,
sonic velocity, and bulk density on the dried samples. Then the
samples were saturated under vacuum and the same measurements were
repeated for the saturated samples. Because the samples were not well
consolidated a number of samples was damaged or destroyed during
saturation, resulting in a lower number of saturated measurements.
Figure~\ref{fig:bw_lab_hist} shows histograms of the results.

\begin{figure}[htbp]
  \centering
  \includegraphics[width=14cm,clip,trim=0 10 0 0]
  {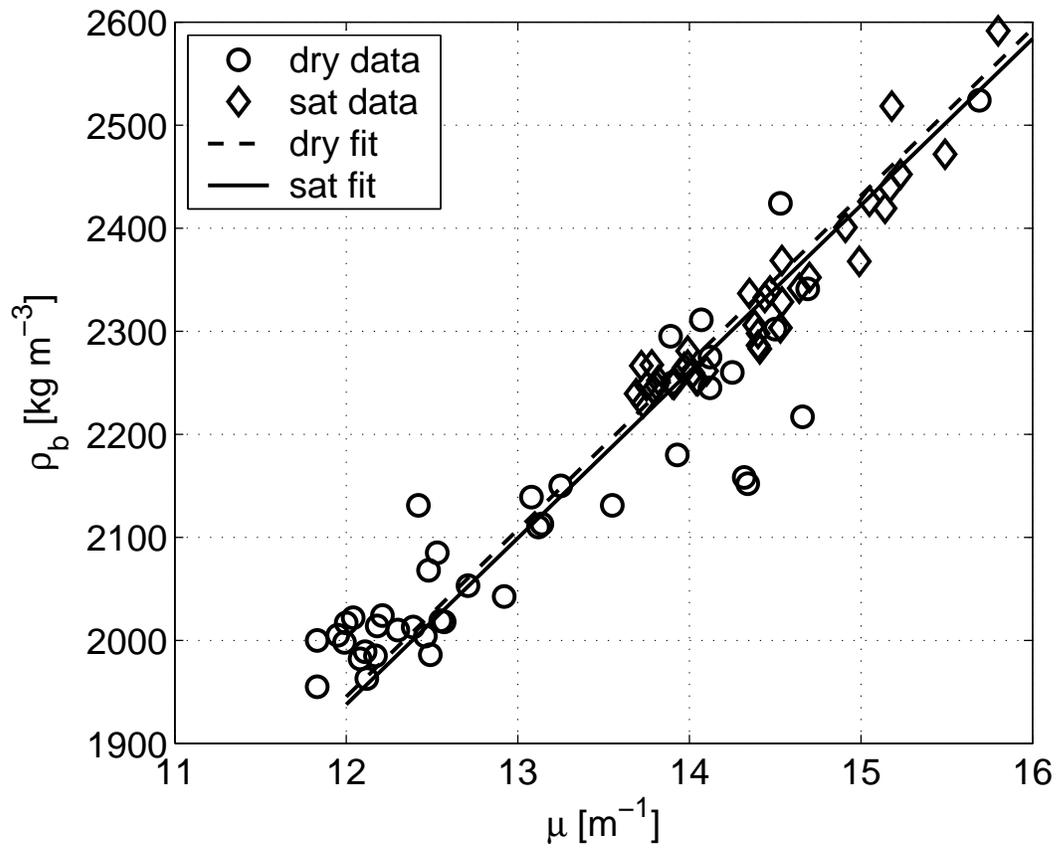} 
  \caption{Calibration of  bulk density measurements. The absorption
    coefficient $\mu$ measured on dry and saturated samples is
    converted into bulk density using
    equation~\ref{eq:lab_rhob_calibration_fit}.}
  \label{fig:bulk_density_calibration}
\end{figure}

\begin{figure}[htbp] 
  \centering
  \includegraphics[width=12cm,clip,trim=0 10 0 0]
  {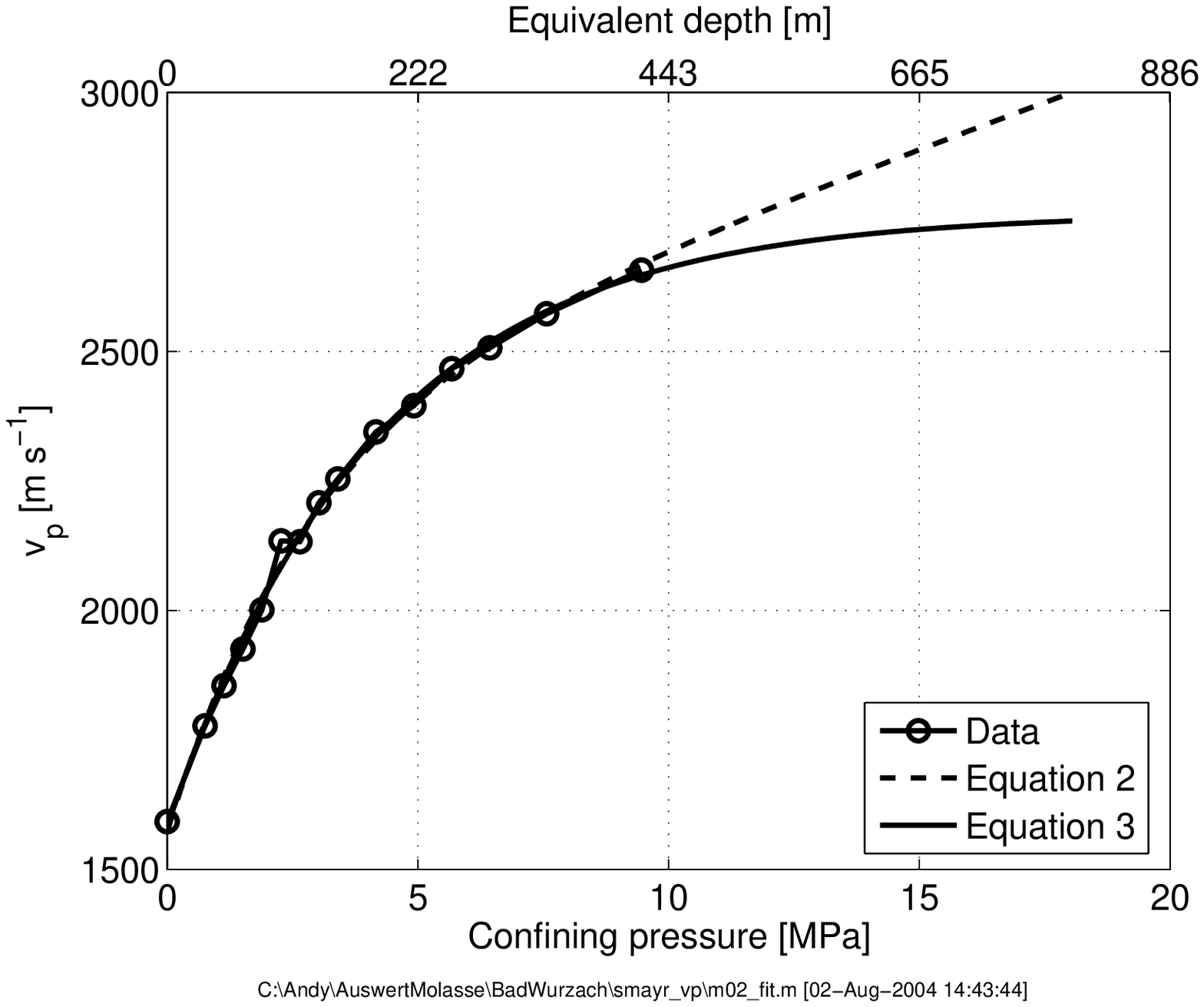}
  \includegraphics[width=12cm,clip,trim=0 10 0 0]
  {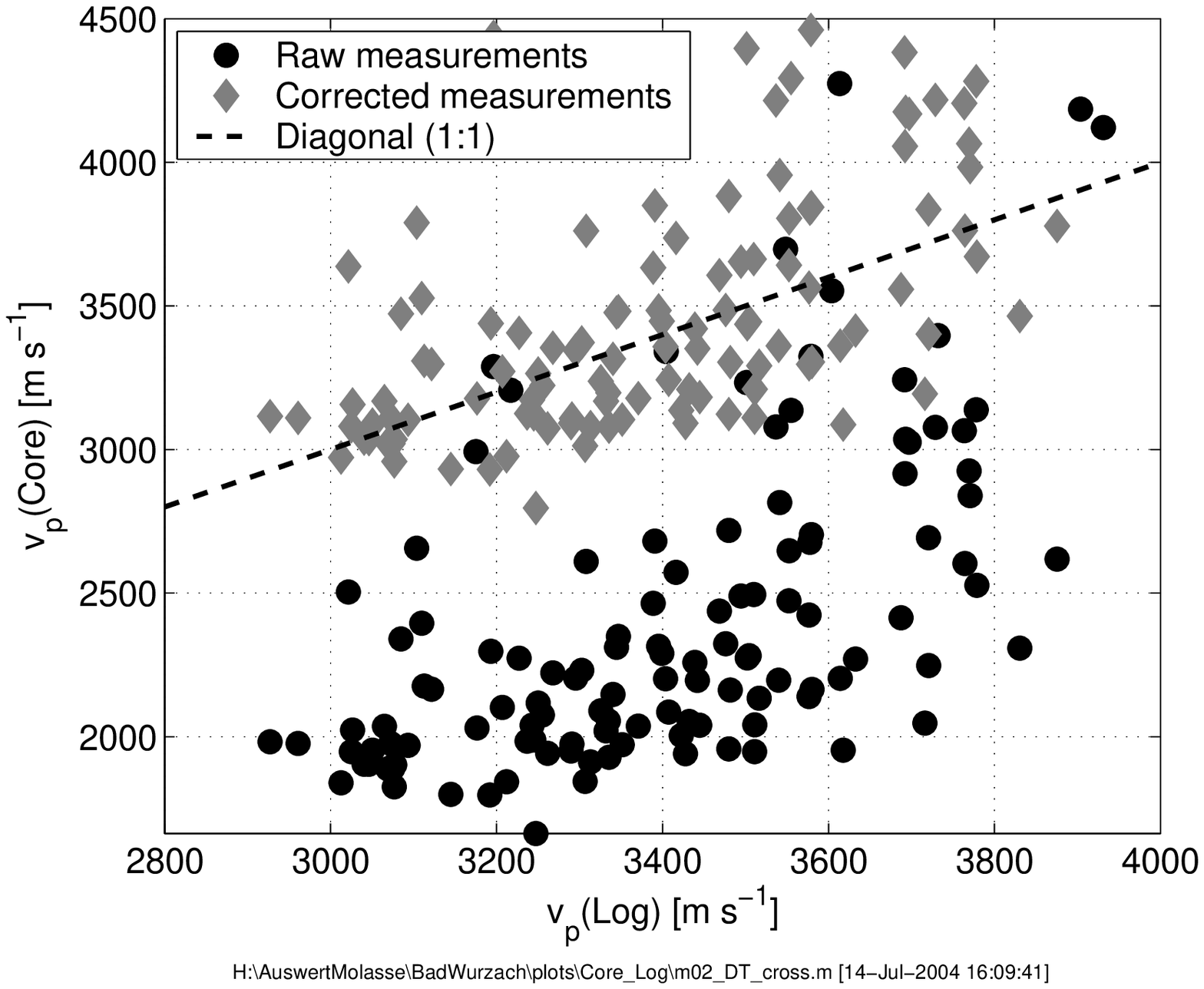}
  \caption{Variation of sonic velocity with confining pressure. Top:
    Laboratory data together with linear-exponential and exponential
    fit. The quality of the fit differs only in the extrapolated
    range. Bottom: Data before and after pressure correction based on
    equation~\protect{\ref{eq:vp_dependence_on_confining_pressure_2}}.}
  \label{fig:bw_vp_pressure_correction}
\end{figure}

Thermal conductivity was measured using the optical scanning technique
which yields a continuous profile of thermal conductivity along the
core axis of the sample \citep{Popov1997,Popov1999,Surma2003}.
  We also measured thermal conductivity perpendicular to the
core axis to obtain anisotropy. Anisotropy ratios were generally less
than 3--4\prc{} so that we assume the rock to be isotropic.  
Values for each sample were computed as averages of the scanning line.
We measured bulk density and sonic velocity with a multi-sensor
core-logger (e.~g.~\cite{Weber1997a}). The measurement is performed
perpendicular to the core axis. Both bulk density and sonic velocity
records need special attention before they can be used in the
analysis.  Bulk density is measured by gamma-ray absorption mainly due
to Compton scattering \cite{Ellis1987}. It is calibrated routinely
using an aluminium standard.  Since electron density is measured
rather than real density, better results are obtained when using a
calibration standard with a similar electron density
\citep{Ellis1987}.  Therefore we measured bulk density and porosity
independently on a number samples using either Archimedes´ method or
gas-state and solid-state pyknometers.  The obtained bulk density is
then plotted versus the absorption coefficient $\mu$ measured with the
core-logger for dry and saturated samples.
Figure~\ref{fig:bulk_density_calibration} shows that a linear
relationship exists which can be used for calibration.  For dry and
saturated measurements the following equations are obtained:
\begin{eqnarray}
  \label{eq:lab_rhob_calibration_fit}
  \rho_{b,s} &=& 161.5 \mu_s\qquad\mathrm{(saturated)} \\
  \rho_{b,d} &=& 162.1 \mu_d\qquad\mathrm{(dry)} \nonumber
\end{eqnarray}
The slope is slightly lower for saturated measurements. A detailed
  analysis shows that saturated gamma density should be
corrected using a separate porosity measurement
\cite{Hearst2000}.  However, considering the data
scatter and the limited range of the saturated measurements this can
be included in the single calibration factor.  Sonic velocity was
measured on the samples under ambient pressure and we found a large
difference in velocity obtained from wireline logs and cores
(Figure~\ref{fig:bw_vp_pressure_correction}, bottom panel). It is well
known that changes in pressure have a profound influence on elastic
properties (e.~g.~\cite{Mayr2002,Zimmerman1986}). An empirical
relationship with linear and exponential terms is usually used to
correct this effect.  In order to quantify the correction, sonic
velocity was measured on a sample under uniaxial pressure as described
in \cite{Mayr2002}. We fitted two exponential equations to this data
(Figure~\ref{fig:bw_vp_pressure_correction}), both with an exponential
term but one without a linear term . The following coefficients were
obtained, with pressure $P$ in MPa and velocity $v$ in \uvel{}:
\begin{eqnarray}
  \label{eq:vp_dependence_on_confining_pressure}
  v(p) &=& 2360 + 787  \e^{0.356P}  + 35.7 P\\ 
  v(p) &=& 2760 + 1180 \e^{0.243P}
  \label{eq:vp_dependence_on_confining_pressure_2} 
\end{eqnarray}
The rms-errors are 19\uvel{}  
(eq.~\ref{eq:vp_dependence_on_confining_pressure}) and 18\uvel{}
(eq.~\ref{eq:vp_dependence_on_confining_pressure_2}). Both equations
fit the pressure-velocity data equally well and differ only in the
extrapolated range from 450\umeter{} to 820\umeter{}.  This is,
however, the depth range of the samples and
equation~\ref{eq:vp_dependence_on_confining_pressure_2} yields better
results when we compare corrected laboratory data to logging data.
Equation~\ref{eq:vp_dependence_on_confining_pressure} slightly
over-corrects the laboratory data. This is probably caused by the fact
that the measurements only cover a range of up to 10\upres{}.  In this
pressure range the exponential term is more significant, and the
linear term is not well defined. Therefore, we used
equation~\ref{eq:vp_dependence_on_confining_pressure_2} to correct the
measurements (Figure~\ref{fig:bw_vp_pressure_correction}, lower part).
The remaining scatter after correction may be due to the use of only
one set of fit parameters. A detailed analysis should take into
account the differing composition of the samples. Because of these
limitations, we did not use sonic velocity in our analysis of
laboratory and logging properties.  We did use it, however, for
correlation analysis of properties measured in the laboratory.
Following this preliminary work of preprocessing laboratory data the
next step is to examine correlations of properties measured in the
laboratory. We discuss the results of this work in the next section.

\section{Laboratory results}
\label{sec:corr_lab_props}

Figure~\ref{fig:bw_lab_hist} shows a summary of the measurements
processed as described in the previous section. Dry properties show a
broader distribution of values, in the case of thermal conductivity it
is even bimodal. This follows from the fact that the ratio of fluid
and matrix properties is larger for the rock/air-system than for the
rock/water-system. Hence, also the range of the effective values is
larger for dry properties.

\begin{figure}[htbp]
  \centering
  \includegraphics[width=14cm,clip,trim=0 20 0 0]
  {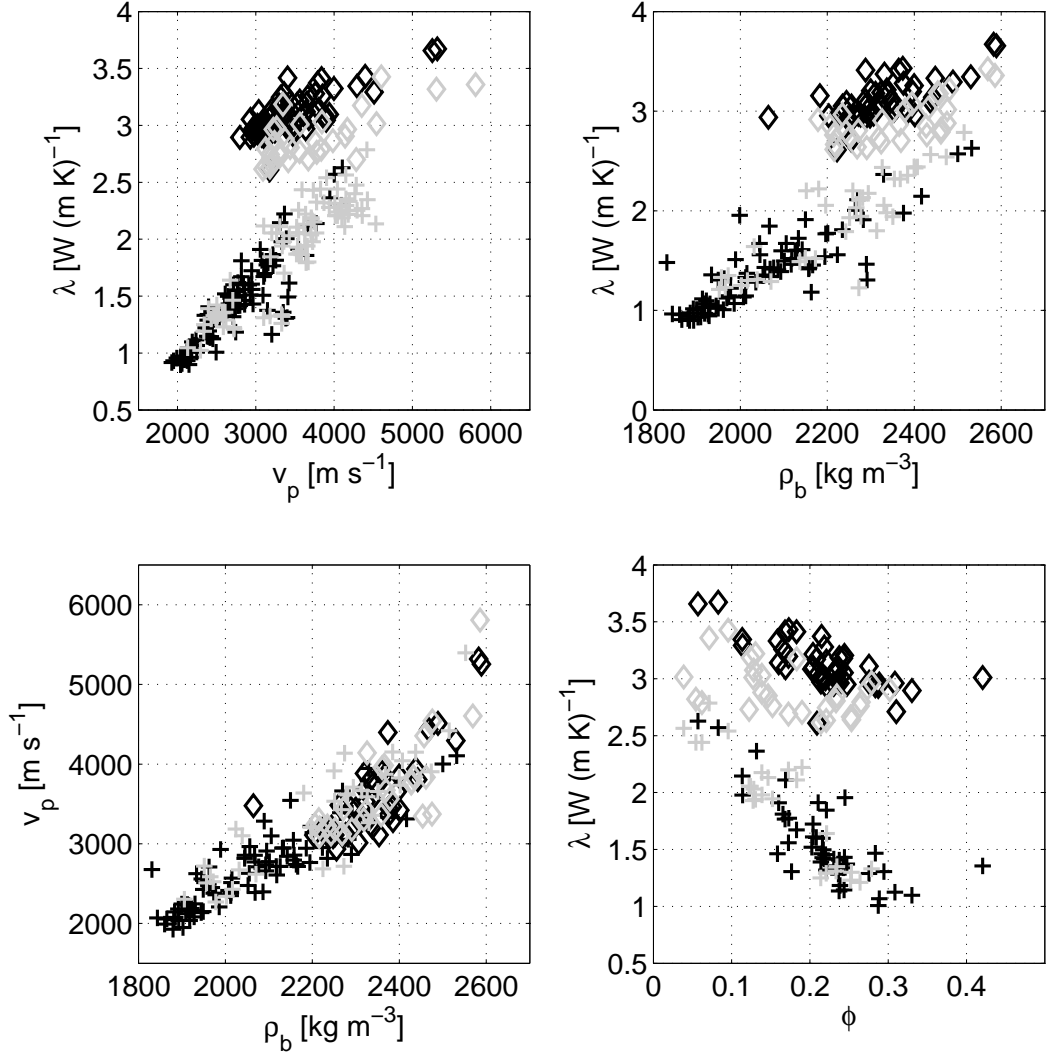}
  \caption{Cross plot of properties measured in the laboratory:
    Thermal conductivity $\lambda$ versus sonic velocity $v_p$ (top
    left), bulk density $\rho_b$ (top right), and porosity $\phi$
    (bottom right). $\phi$ is derived from dry and saturated gamma
    density measurements. Sonic velocity versus bulk density is
    plotted in the lower left. Black and grey symbols correspond to
    sandy and carbonaceous samples. Dry and saturated measurements are
    shown as (+) and ($\diamondsuit$), respectively.}
  \label{fig:bw_lab_crossplot}
\end{figure}

Our goal in measuring the various properties is to predict thermal
conductivity from other petrophysical properties which can be measured
in-situ. For this purpose we analysed linear correlations between
thermal conductivity in dry and saturated condition and and other
measurements in their corresponding saturation state.
Figure~\ref{fig:bw_lab_crossplot} shows cross-plots of these data.
  Porosity was computed from dry and saturated bulk density.
We did not use sonic measurements to compute porosity as
velocity-porosity models work best when velocity is measured under
confining pressure such that  the rock is at its terminal
velocity \cite{Wyllie1958,Mavko1998}.  We grouped the samples
according to their lithology:  The first group contains shaly
sandstone samples (\lq{}\lq{}sandy samples\rq{}\rq{}), the second
group holds marlstones (\lq\lq carbonaceous samples\rq{}\rq{}). This
division is somewhat preliminary as it is based on geological core
descriptions and not on an analysis of the mineral assemblages.

A regression analysis of thermal conductivity versus sonic velocity,
bulk density and porosity, respectively, was performed.  For each
combination a linear equation of the form
\begin{equation}
  \label{eq:linear_fit}
  y=a_1 x + a_0  
\end{equation}
was fitted to the data.  The regression analysis uses a  
total-least-squares solution to account for measurement errors both in
the dependent and independent variables \cite{VanHuffel1991}. The
Jackknife-method yields the variances of the parameter estimates
\cite{Shao1995_jackknife}.  We also computed correlation coefficient
$R$ and rms-error to assess the quality of the fit \cite{Blobel1998}.

Table~\ref{tab:bw_therm_con_pred_results} summarises the results.  If
the rms-error is interpreted as the predictive error of the computed
relationships it can be deduced that it is possible to calculate
thermal conductivity from density and or sonic velocity with an
average accuracy of about 0.1--0.2\uwlf{}, given an appropriate data
set for calibration. Correlation coefficients are largest for dry
properties. This is due to the larger contrast in rock and pore volume
properties for dry samples. Thus, porosity variations cause larger
variations in the effective properties of the two-component system and
therefore stronger correlations for dry samples. On the other hand, the
low correlation coefficient between thermal conductivity and porosity
is due to the complex relationship between these properties.  Whereas
a linear equation will be sufficient to describe the relationship
between thermal conductivity and other petrophysical properties, it is
inadequate for porosity. This is addressed in more detail in section
\ref{sec:choice_of_mixing_law}. Grouping samples according to their
lithology has a larger effect on the quality of the fit for saturated
than for dry samples.  Figure~\ref{fig:bw_lab_crossplot} shows in the
crossplot of thermal conductivity versus sonic velocity or bulk
density that the two lithology groups have the same general trend for
dry samples but are separated for saturated measurements.  This again
is an effect of the strong influence of porosity for dry samples which
masks variations due to lithology. From our analysis it thus appears
that dry measurements are most sensitive to porosity changes whereas
saturated ones reflect both variations in lithology as well as
porosity.

The question arises if a correlation using more than one petrophysical
property improvea the prediction of thermal conductivity.
Table~\ref{tab:bw_multi_regression} shows the results of a multiple
linear regression for thermal conductivity. The equation has the
form:
\begin{equation}
  \label{eq:multi_linear_regression}
  \lambda = a_0 + a_1 v_p + a_2 \rho + a_3 \phi
\end{equation}
There is only slight improvement over the simple regressions. In
practise this improvement might be outweighed by the additional
effort for performing the measurements.  If the samples are
considered to be composed of the three components sand, shale, and
carbonate, the quality of the predicted thermal conductivity is
directly related to the quality of prediction of these three
components from the measured properties.  However, quartz and calcite
differ in density and velocity by only 3 and 13~\%, respectively.
Shales have a large variability in their physical properties, but they
usually differ considerably from quartz and calcite
\cite{Ellis1987,Crain1986}.  Therefore, any of the measurements will
be more sensitive to variations of shale content and porosity than to
changes in carbonate and sand content. Thus, a combination of sonic
velocity and bulk density does not provide significantly more
information than each of them alone.  This fact can be assessed in a
cross-plot of these two properties (Figure~\ref{fig:bw_lab_crossplot},
lower left panel).  Sandy and carbonaceous samples essentially plot on
top of each other.  Additional measurements, such as natural gamma
radiation for instance, are required to characterise our samples
better. The contrast in thermal conductivity between quartz and
calcite is about 60~\%{}, sufficiently large to separate them in a
cross-plot (Figure~\ref{fig:bw_lab_crossplot}, upper left). This
offers opportunities for characterising lithology, as conductivity can
be determined rapidly and continuously along a core.

\section{Choosing an appropriate mixing law}
\label{sec:choice_of_mixing_law}

\begin{figure}[htbp]
  \centering
  \includegraphics[width=14cm,clip,trim=0 10 0 0]
  {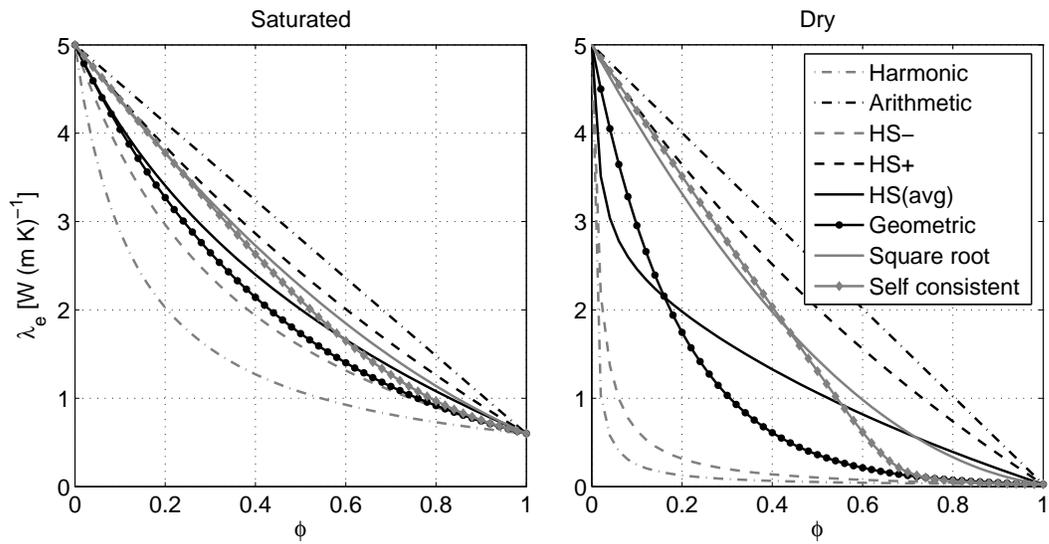}
  \caption{Comparison of different mixing laws for thermal
    conductivity: HS\up{$\pm$} - Hashin-Shtrikman upper and lower
    bounds.  HS(avg) - mean of HS\up{+ }and HS\up{--} bounds. Results
    are shown for saturated and dry rock samples. Matrix, water, and
    air thermal conductivities are taken to be 5\uwlf{}, 0.6\uwlf{},
    and 0.026\uwlf{}, respectively. Choosing an inappropriate mixing
    law results in much larger errors for dry samples.}
  \label{fig:mixing_laws_tc_without_aspect}
\end{figure}

A mixing law for calculation of the effective thermal conductivity of
a composite medium according to the content and thermal conductivity
of its components is required both in the analysis of thermal
conductivity measured in the laboratory on dry and saturated samples
as well as in the prediction of thermal conductivity from logging
data. Several models have been proposed and it is instructive to
review the relationships most commonly used in geothermics
\cite{Beardsmore2001}.  Effective thermal conductivity $\lambda$ of a
layered medium with thermal conductivities $\lambda_1$ and $\lambda_2$
depends on the direction of the temperature gradient. If heat flow is
parallel to the layering the effective conductivity is equal to the
arithmetic mean ($\lambda_a$) layer thermal conductivities, weighted
by their volume fractions. For perpendicular heat flow it corresponds
to the harmonic mean ($\lambda_h$) \cite{Carslaw1959}:
\begin{eqnarray}
  \label{eq:arithmetic_and_harmonic_mean}
  \lambda_a &=& V_1 \lambda_1 + V_2 \lambda_2 \\
  \lambda_h &=& \frac{V_1}{\lambda_1}  + \frac{V_2}{\lambda_2}
\end{eqnarray}
While these averages are useful to estimate the average thermal
conductivity of a vertical rock sequence they are inappropriate for
estimating effective sample thermal conductivities. Narrower bounds
can be derived by assuming a geometry where the solid consists of
spheres dispersed in the pore fluid or were the fluid is confined in
spherical inclusions in the rock matrix \cite{Hashin1962,Horai1971}.
This configurations yield the lower and upper Hashin-Shtrikman
(HS\up{--}, HS\up{+}) bounds, respectively:
\begin{eqnarray}
  \label{eq:upper_and_lower_hashi_shtrikman_bounds}
  \lambda_{HS^-} &=& \lambda_p + \frac{1-\phi}{
    \frac{1}{\lambda_m-\lambda_p}+\frac{\phi}{3\lambda_p}} \\ 
  \lambda_{HS^+} &=& \lambda_m + \frac{\phi}{
    \frac{1}{\lambda_p-\lambda_m}+\frac{\phi}{3\lambda_m}}
\end{eqnarray}
Here $\phi$ denotes porosity, and the subscripts $p,m$ denote pore and
rock matrix properties, respectively.  These bounds are of theoretical
importance because effective thermal conductivities of rock samples
should generally fall in between these bounds. However, in many cases
they are to far apart to be of practical use. In this situation an
estimate can only be obtained from empirical relationships such as the
geometric mean $\lambda_g$ \cite{Sass1971a}, that is often used in
geothermal studies:
\begin{equation}
  \label{eq:geometric_average}
  \lambda_g = \lambda_p^{\phi}  \lambda_m^{1-\phi}
\end{equation}
However, other researcher preferred to use the average of the upper
and lower HS bounds \cite{Horai1971}, or the square root average
$\lambda_s$ \cite{Roy_1981_thermophysical_properties,Beardsmore2001}:
\begin{equation}
  \label{eq:square_root_average}
  \sqrt{\lambda_s} = \phi\sqrt{\lambda_p} + (1-\phi)\sqrt{\lambda_m} 
\end{equation}
The self consistent approach \cite{Hill1965,Budiansky_1970_composites}
is popular for elastic properties but is not widely used in geothermal
research. For porosities typical for rocks it gives results similar to
the square-root average. The thermal conductivity $\lambda_b$ for a
two component medium is given by the equation:  
\begin{equation}
  \label{eq:budiansky_self_consistent}
  \frac{\phi}{2/3 + \lambda_p/(3\lambda_b)} + \frac{1-\phi}{2/3 +
    \lambda_m/(3\lambda_b)} = 1
\end{equation}
  The particular choice of a model becomes important when the
contrast in thermal conductivity of the constituents increases. 
Figure~\ref{fig:mixing_laws_tc_without_aspect} illustrates this by
showing bounds and estimates for the saturated and the dry case.
Thermal conductivity is assumed to be 5\uwlf{} for the matrix and
0.6\uwlf{} and 0.026\uwlf{} for water and air, respectively. In this
case the maximum difference between lower and upper HS bounds are 3.7
and 1.0\uwlf{} for dry and saturated samples. For a two-phase mineral
assemblage of crystalline quartz (7.7\uwlf{}) and orthoclase
(2.3\uwlf{}) the maximum difference is as low as 0.4\uwlf{}.  This
shows that the choice of a correct mixing law for a mineral assemblage
is somewhat arbitrary while it is essential when considering dry
samples.  Although a dry sample will rarely occur in nature, these
considerations are important when laboratory measurements on dry
samples are used to predict in-situ saturated thermal conductivity.

Another important conclusion is, that in the case of a mineral
assemblage the geometric mixing law closely follows the lower
Hashin-Shtrikman bound, corresponding to a rock model consisting of
grains suspended in a fluid.  The square root law, on the other hand,
is very close to the upper Hashin-Shtrikman bound and could be
interpreted as a well lithified rock with spherical pores. Thus, as
each particular empirical mixing laws corresponds to a particular
rock-structure, one single model cannot be adequate for all different
rock structures.

Additional parameters can be introduced in order to incorporate rock
structure into a mixing law.  Several models assume spheroidal pores
where $\alpha$ is the aspect ratio of the spheroids
\cite{Schulz1981,Zimmerman1989,Buntebarth1998,Popov2003}.    In
an application of the model by Zimmerman \cite{Zimmerman1989} aspect
ratios as low as 0.1 were found for basalts \cite{Horai1991}, much
less than the actual aspect of the pores. It was regarded as a value
representing the aspect ratio of the grain contact rather than that of
the pores.
%A similar interpretation
%is that such a low value for the aspect ratio yields a relationship
%close to the lower HS bound. This suggests a suspension of grains
%rather than a solid rock matrix with pore space.

From this discussion it appears that mixing laws of varying complexity
can be used in different situations: Thermal conductivity of a mineral
assemblage or a water-saturated rock can be computed with sufficient
accuracy using a geometric mixing law. For measurement on dry samples
the pore structure needs to be taken into account. The model proposed
in \cite{Zimmerman1989} was therefore applied to our laboratory data
set in order to determine the aspect ratio $\alpha$.  The model
assumes a homogeneous mixture of randomly distributed spheroids.  For
a rock with oblate spheroidal pores one obtains the effective thermal
conductivity $\lambda$:
\begin{equation}
\label{eq:zimmerman_1989_model}
  \frac{\lambda }{{\lambda _m }} = \frac{{(1 - \phi )(1 - r) + r\beta
      \phi }}{{(1 - \phi )(1 - r) + \beta \phi }}
\end{equation}
The parameters $\beta$, $M$, and $\theta$ are defined by:
 
\begin{eqnarray}
  \label{eq:zimmerman_1989_explanations}
  r &=& \frac{{\lambda _f }}{{\lambda _m }}\\  
  \beta  &=& \frac{{1 - r}}{3}  \left( {\frac{4}{{2 + (r -
          1)M}} + \frac{1}{{1+(r - 1)(1 - M)}}} \right) \\
  M &=& \frac{{2\theta  - \sin 2\theta }}{{2\tan \theta \sin ^2 \theta
    }} \\
  \label{eq:zimmerman_1989_explanations_4}
  \theta  &=& \arccos \alpha 
\end{eqnarray} 
 
\begin{figure}[htbp]
  \centering
  \includegraphics[width=14cm,clip,trim=0 10 0 0]
  {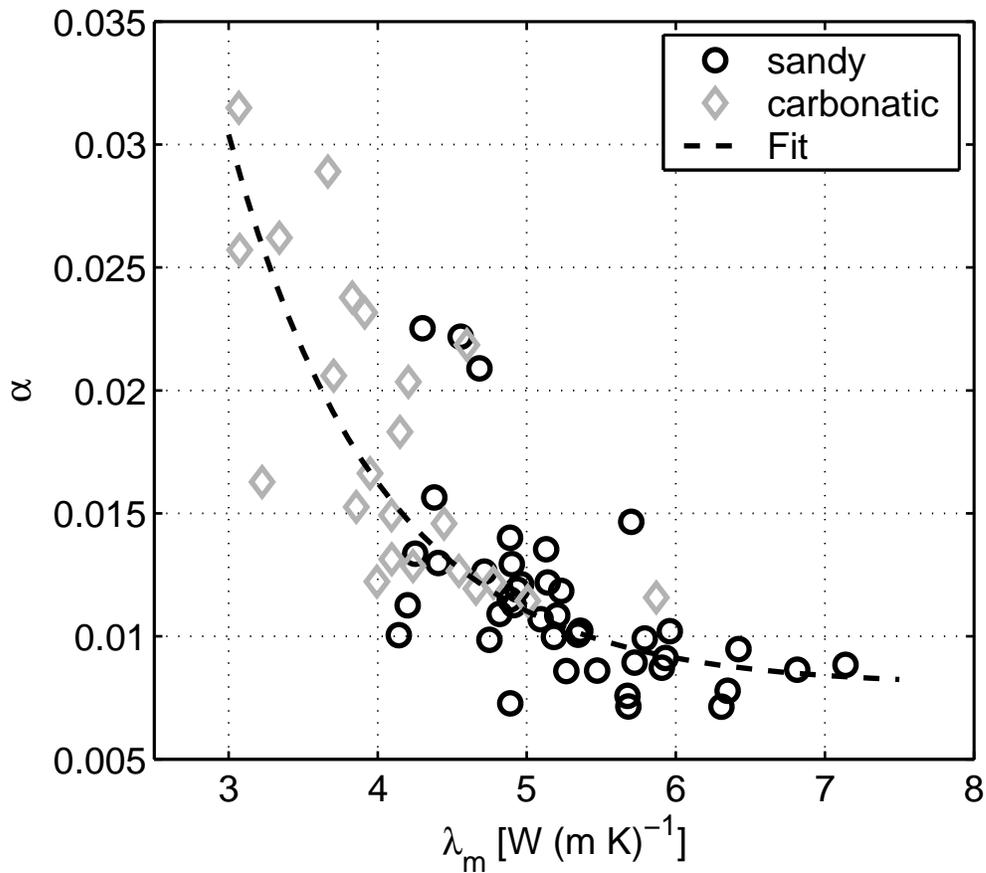}
  \caption{Variation of aspect ratio $\alpha$ versus matrix thermal
    conductivity $\lambda_m$. Both properties are computed using
    equations~\ref{eq:zimmerman_1989_model} and
    \ref{eq:zimmerman_1989_explanations}. We used dry and saturated
    thermal conductivity and porosity derived from bulk density as
    input to the model.}
  \label{fig:zimm_aspect_ratios}
\end{figure}

  Again, $\alpha$ is the aspect ratio of the
spheroidal  inclusions. The model described by equations
\ref{eq:zimmerman_1989_model} --
\ref{eq:zimmerman_1989_explanations_4} consists of five parameters:
effective, matrix, and pore fluid thermal conductivity, porosity, and
aspect ratio. Our measurements of saturated and dry thermal
conductivity yield two equations of this type which differ only in the
effective thermal conductivity $\lambda$ and the pore fluid thermal
conductivity $\lambda_f$. This yields seven parameters in total. If we
use the two thermal conductivity measurements, known values of water
and air thermal conductivity, and an independent porosity measurement,
we obtain two equations with two unknowns: Matrix thermal conductivity
and pore aspect ratio. We solved the system for our data by a
nonlinear iterative algorithm.   
Figure~\ref{fig:zimm_aspect_ratios} shows a crossplot of the computed
values of matrix thermal conductivity $\lambda_m$ and aspect ratio
$\alpha$. The median values of $\alpha$ are 0.011 and 0.016 for sandy
and carbonaceous samples, respectively.  This is about one magnitude
less than the values reported in \cite{Horai1991}.  However, those
values were measured on igneous rocks and crack aspect ratios based on
sonic or mechanical measurements generally range from
10\up{-2}--10\up{-3} \cite{Cheng1979,Zimmerman1984}. Median matrix
thermal conductivities are 5.14\uwlf{} and 4.09\uwlf{} for sandy and
carbonaceous samples.

\section{Thermal conductivity predicted from wireline data}
\label{sec:tc_pred_using_han86}

\begin{figure}[htbp]
  \centering
  \includegraphics[width=7cm]{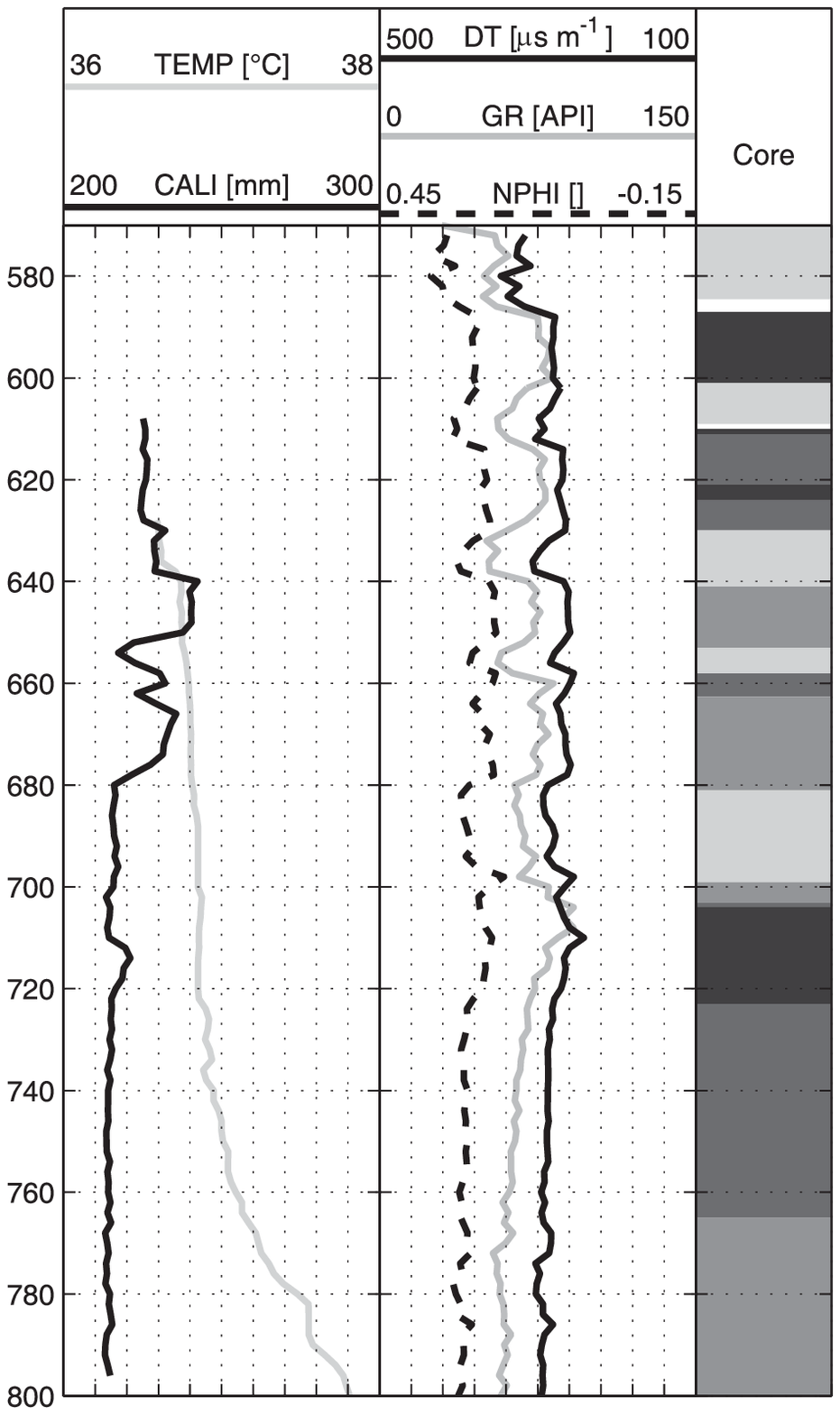}
  \includegraphics[width=4.5cm]{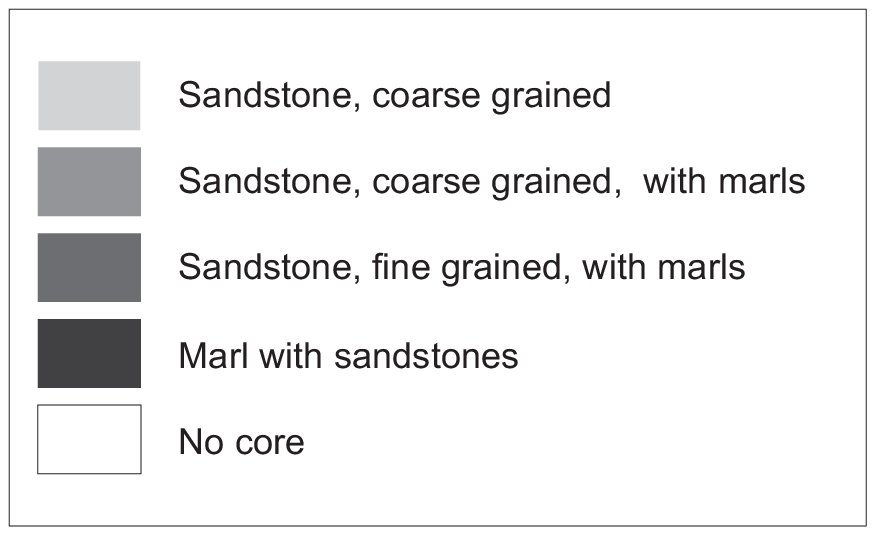}
  \caption{Composite log showing caliper, sonic, neutron porosity,
    gamma-ray, and core lithology.} 
  \label{fig:bw_composite_qc}
\end{figure}

Wireline logs of natural   gamma radiation (GR), neutron
porosity (NPHI), and bulk density (RHOB) were available for analysis
in the depth range 570--800\umeter{}. In the depth range from
600--800\umeter{} caliper (CALI) and temperature (TEMP) had been
logged during an aquifer test. The logs are shown together with the
core lithology in figure~\ref{fig:bw_composite_qc}. The temperature
log is strongly disturbed by water flowing from the formation with
temperatures varying from 35--37°C. Unfortunately the disturbance
renders a quantitative interpretation of temperatures with regard to the
conductive regime impossible. The hole was drilled with 8.5\uinch{}
diameter and the caliper generally reads below 9\uinch{}. Some larger
diameter sections (630--680\umeter{}) are consistent with lithological
changes. The core lithology showed no signs of hydrocarbons and the
drilled sequence is not known as a hydrocarbon reservoir. Thus, log
data quality can be regarded sufficient for quantitative
interpretation.

We applied several editing steps before we analysed the logging data.
For a general discussion of these steps see for example
\cite{Jun2002,Serra1984,Ellis1987}. Logging curves from different tool
runs were corrected to common depth points, bad data were eliminated,
and environmental corrections applied. Core depths are shifted to
match logging depth.    Core data are smoothed before
they are compared to log curves since these have a lower depth
resolution. For this purpose we used an inverse distance algorithm
that employs an Gaussian weighting function:
\begin{eqnarray}
  \label{eq:gauss_weight_1}
  w_k &=& \sum_{i=1}^{N} \e^{-
    \left(
     \frac{z_k-z_i}{r_w} 
    \right)^2} \\
  v_k &=& \frac{1}{w_k} \sum_{i=1}^{N} v_i \e^{-
    \left(
     \frac{z_k-z_i}{r_w} 
    \right)^2}
\end{eqnarray}
Here $v_k$ are data values, $z_i$ is the depth of the data value, and
$w_k$ is the normalisation constant. We took the weighting distance
$r_w$ to be 0.5\umeter{}. 

In general the response $R^j$ of a wireline tool $j$ is determined by
the volume fractions $V_i$ of the rock components $i$ and their
theoretical log response $T_i^j$.  Assuming a linear relationship this
results in
\begin{equation}
  \label{eq:log_response}
  R^j = \sum_{i} V_i T_i^j,
\end{equation}
with the constraint that $\sum_i V_i = 1$
\cite{Doveton1979,Hoppie_1996_lithological_characterization}.  This
equation is correct for properties like bulk density or neutron
porosity. For acoustic properties, on the other hand, it is not
generally valid but results in empirical relationships. If for
instance the slowness is considered as the sonic measurement,
equation~\ref{eq:log_response} implies Wyllie´s classical travel-time
average \cite{Wyllie1956}. If the number of constituents equals the
number of tool responses equation~\ref{eq:log_response} has one
solution. If the number of equations is larger than the number of
components, the system is overdetermined and
equation~\ref{eq:log_response} can be solved in a least-squares sense.
Thus, lithological composition can be computed from a sufficient
number of logs. Based on this composition and the known values of
thermal conductivity of the components, an effective thermal
conductivity can be computed. The geometric average law is used to
compute the effective thermal conductivity of the mineral mixture.
This value will be used as the matrix thermal conductivity.  The
effective conductivity of dry and water saturated rocks is then
calculated from equation~\ref{eq:zimmerman_1989_model} using a median
aspect ratio of 0.012. As we want to compare the results to our
laboratory data, no temperature correction is necessary at this stage.

As discussed previously for laboratory data, we explore different
models with varying simplicity to evaluate how well thermal
conductivity can be described by logging data under different
circumstances. Two models are of particular interest: (1) One model
consists of a mixture of sand, shale, and carbonate and uses the
information of all logs available for the borehole; (2) Another model
consists only of the two components sand and shale, ignores the
carbonate fraction, and uses only wireline logs of slowness and
natural gamma radiation, DT and GR, respectively. This model is
required for older wells where only these two logs are available.
Although these wells have not been logged with modern tools they
comprise a large fraction of the available dataset.

\begin{figure}[htbp]
  \centering
  \includegraphics[width=15cm,clip,trim=0 10 0 0]
  {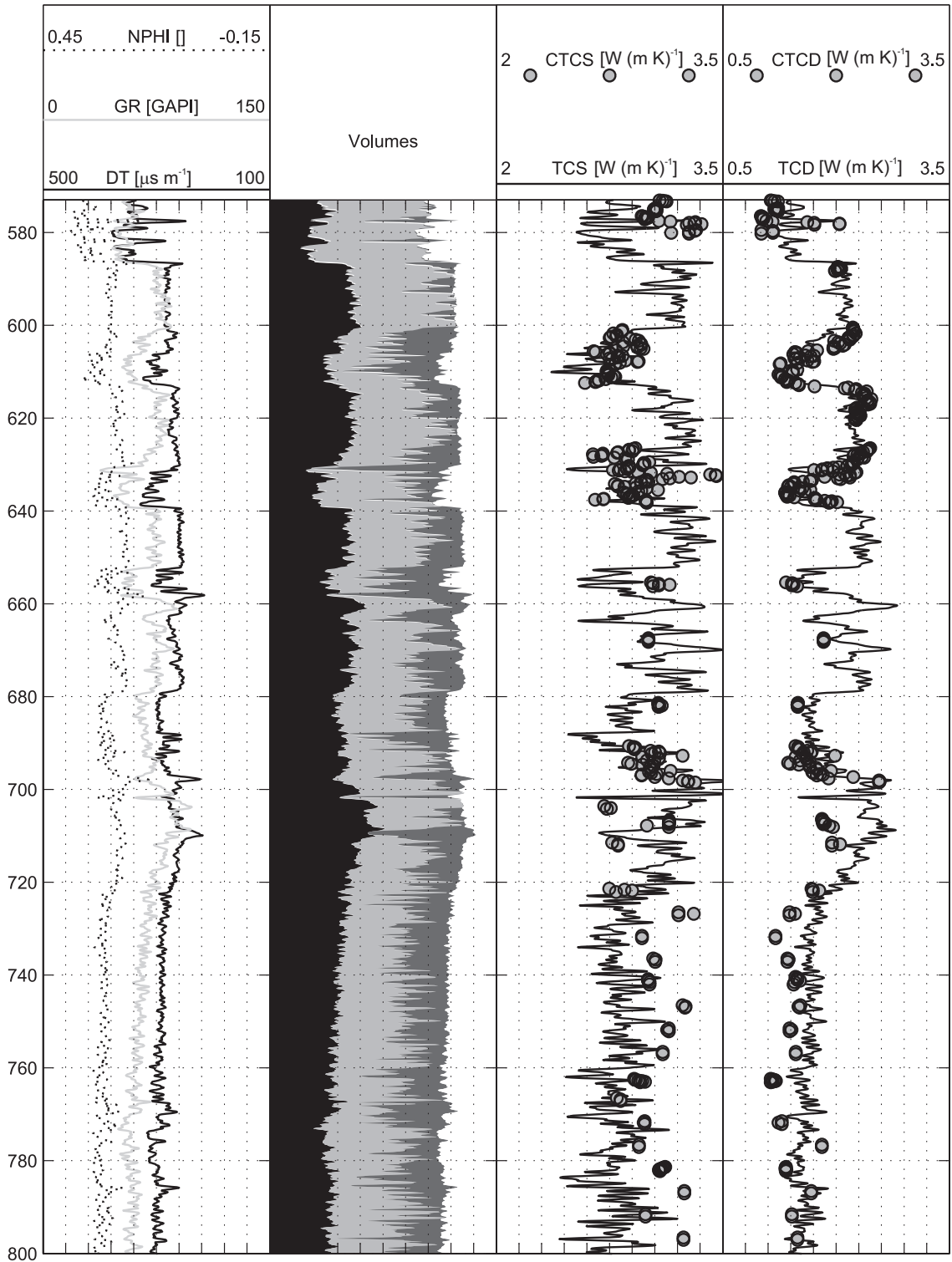}
  \caption{Sand-Shale-Carbonate model for the borehole studied. Input data
    (left panel): NPHI - neutron porosity; GR - natural gamma
    radiation; DT - acoustic slowness. Input logs are used to compute
    the composition (middle left panel). Colour coding: black - shale;
    light grey - sand; dark grey - carbonates; white - porosity.
    Composition is then used to compute saturated and dry thermal
    conductivity, TCS and TCD, respectively (middle right and right
    panel). Thermal conductivity derived from logs is compared to
    saturated and dry core data, CTCS and CTCD.}
  \label{fig:bw_full_model_log_panel}
\end{figure}

The full model uses logs of gamma density (RHOB), natural gamma
radiation (GR), and slowness (DT) as input logs.  We employed a
commercial software package specialised in deriving rock composition
from wireline logs \cite{ELANPlus1999}. The three components sand,
shale, and carbonate were parameterised in the inversion using
response values of the minerals quartz, glauconite, and calcite.  The
choice of the shale mineral glauconite, an iron-rich variety of
illite, is based on the geological description of the cores and
general information about the geology of the Upper Freshwater Molasse
\cite{Geyer1991}. It might not be the only shale mineral present, but
the values for illite, another abundant shale mineral, are very
similar to the ones we used. Values for the logging properties and
mineral thermal conductivities are summarised in
Table~\ref{tab:bw_param_summary_for_tc_predicitive_models}. Thermal
conductivity values for quartz and calcite are 7.69\uwlf{} and
3.59\uwlf{}, respectively \cite{Clauser1995a}. The choice for the
shale fraction is more difficult. Glauconite was measured with a value
of 1.6\uwlf{} \cite{Horai1971}. We found that an optimal fit can be
obtained with a higher value of 2.2\uwlf{}. This discrepancy could be
easily explained by the variability of properties for shale minerals,
but could also indicate that small amounts of other minerals are
present which are not accounted for. There is generally a good
agreement between computed and measured thermal conductivity
(Figure~\ref{fig:bw_full_model_log_panel}). The rms error of the
reconstruction is 0.27\uwlf{} and 0.28\uwlf{} for saturated and dry
thermal conductivity. Although the rms-error is slightly larger for
dry properties, they can be reproduced much better than saturated
properties because the large conductivity contrast of air/rock matrix
enhances the variations in thermal conductivity. 

\begin{figure}[htbp]
  \centering
  \includegraphics[width=15cm,clip,trim=0 10 0 0]
  {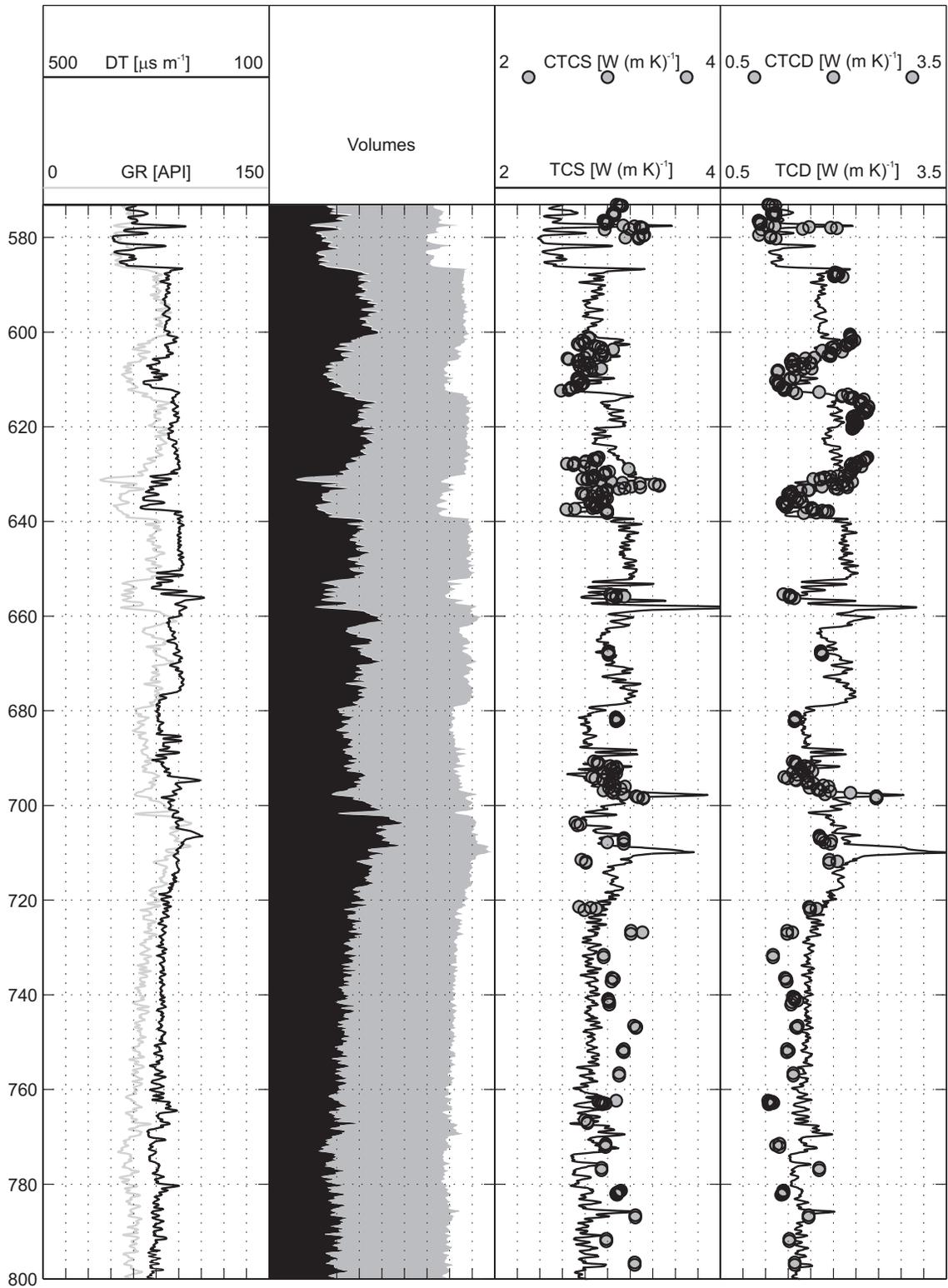}
  \caption{Sand-Shale model for the borehole studied. Input data
    (left panel): GR - natural gamma radiation; DT - acoustic
    slowness. Input logs are used to compute the composition (middle
    left panel). Colour coding: black - shale; light grey - sand; white
    - porosity.  Composition is then used to compute saturated and dry
    thermal conductivity, TCS and TCD, respectively (middle right and
    right panel). Thermal conductivity derived from logs is compared
    to saturated and dry core data, CTCS and CTCD.}
  \label{fig:bw_san_shal_model_log_panel}
\end{figure}

The second predictive model employs a simple mineralogy consisting
only of sand and shale. Additionally, an empirical relationship is
used between sonic velocity $v_p$, shale content $V_{Shale}$, and
porosity $\phi$, established for shaly sandstones and a pressure range
from 10--40\upres{} \cite{Han1986b}. At a pressure of 20\upres{} (about 900\umeter{})
the equation for the sonic velocity is given by
\begin{equation}
  \label{eq:han1986}
  v_p \equnit{[km s\up{-1}]}= 5.49 - 6.94 \phi - 2.17
  V_{Shale}.
\end{equation}
The empirical parameters in this equation are slowly varying functions
of pressure.  For the gamma-ray log we use the linear log response
equation~\ref{eq:log_response} assuming no radiation for the fluid:
\begin{equation}
  \label{eq:gamma_ray_from_composition}
  \li{GR} = \li{GR}_{sand} V_{Sand} + \li{GR}_{Shale} V_{Shale}
\end{equation}
The parameters used for the model are given in
Table~\ref{tab:bw_param_summary_for_tc_predicitive_models}. Using the
mean density of our laboratory measurements of 2330\udens{}, depths
were converted to lithostatic pressure which is needed for
equation~\ref{eq:han1986}. The resulting lithology profile is shown in
figure~\ref{fig:bw_san_shal_model_log_panel}. 

Again we test this model for its ability to predict thermal
conductivities. For this purpose we sampled the compositional log at
the depths of the core measurements. Computed composition and measured
values of thermal conductivity were then used in an inversion to find
optimal mineral thermal conductivities of the constituents
(Table~\ref{tab:bw_param_summary_for_tc_predicitive_models}). A
synthetic log of thermal conductivity is then computed and compared to
the measurements. The rms-misfit of this two-component model is
0.28\uwlf{}, essentially the same as the misfit of the three-component
model.

When comparing the lithology logs for the two models it is apparent
that the volume fraction of the shale component has not changed very
much. The sand fraction of the two-component model also includes the
missing carbonate fraction. This is also reflected in the lower
mineral thermal conductivity of the sand/shale model. As a consequence
the model will be only successful when the ratio of sand and
carbonates does not change too much. This can be seen in
figure~\ref{fig:bw_san_shal_model_log_panel} at depths around
630\umeter{}, where high thermal conductivities cannot be accurately
reproduced by the model.

   Both models display a better fit in the upper part
of the profile than in the lower part below 700\umeter{}. We believe
that this is due to an overestimation of porosity for the cores we
analysed. Average porosity in the depth range 750--800\umeter{} is
20\prc{} for core data but only 17\prc{} for log data. The difference
can be attributed to the release of overburden pressure during coring
\cite{Holt2003_porosity_correction}. However, at this point no
measurements under confining pressure could be performed to verify
this effect.  Depending on the geological setting and method to
compute an overburden correction, correction factors of 0.85--0.95
have been reported \cite{Nieto1994}. It is to be
expected that our samples respond strongly to the pressure relief due
to their poor consolidation. In this situation log derived porosities
might be more accurate and reliable than core derived porosities
adding to the usefulness of log derived thermal conductivity.

  So far, conductivities were computed only for room
temperature.  For  mod\-el\-ing of temperatures in-situ it is
necessary to correct the effect of temperature on thermal
conductivity. Temperature dependent thermal conductivity was measured
for 23~samples of the Molasse Basin, one of them from the borehole
analysed here~\cite{Clauser2002}. An empirical function based
on \cite{Sass1992} of the form
\begin{equation}
  \label{eq:sass_1992}
  \lambda(T) = \frac{\lambda_0}{a + T (b - c/\lambda_0)}
\end{equation}
was used fit the data \cite{Vosteen2003}. Here
$\lambda_0$ is the thermal conductivity at room temperature (25°C).
Coefficients are $a=0.99$, $b=3.4\cdot10^{-3}$\NumUnitDist{}K\up{-1},
and $c=3.9\cdot10^{-3}$\uwlf{}. Using the aquifer temperature of 36°C
at 800\umeter{} depth and a range of 2.5--3.5\uwlf{} for saturated
thermal conductivity the necessary correction amounts to 5--7\prc{}.

\section{Summary and Conclusion}
\label{sec:conclusion}

Empirical relationships between thermal conductivity and other
petrophysical properties depend on local conditions, in particular the
type of diagenesis for the rocks. We analysed core and log data from
one borehole in the Molasse Basin in order to establish a set of such
equations. Empirical relationships were derived from sonic velocity or
bulk density laboratory data that allow to predict thermal
conductivity to an accuracy of about 0.2\uwlf{}, on average.
Predicting thermal conductivity from logging data has a larger error
of about 0.3\uwlf{}. On the one hand, this may be expected in view of
the different spatial resolutions and the problems encountered in the
matching of core and log depths. On the other hand, this is
outweighed by the larger number of sampling points. An important
restriction, as with other studies of this type, is that the results
are restricted to the particular conditions in a specific basin.

An ideal combination of wireline logs for an optimum determination of
lithology would comprise the full suite of nuclear measurements. In
case of the Molasse Basin, we demonstrate that thermal conductivity can
be derived even though the number of available logs is less than
ideal. This is an important result, as for many old wells - drilled
before the advent of modern logging tools - often only a natural gamma
log is available. The fact that thermal conductivities can be
reconstructed from these data alone makes these older wells attractive
for such an analysis.

The discussion of laboratory data and the two compositional models
shows that this log combination is most useful in a setting dominated
by sand and shale fractions. In contrast, a combination of sandstone
and carbonate cannot be well characterised. Fortunately though, these
two components differ by a large contrast in thermal conductivity.
This measurement can be performed rapidly and yields a continuous
profile along the core. Also, thermal conductivity is closely linked
to other petrophysical properties, such as permeability
\cite{Popov2003}. This can be particularly useful, for instance, in
reservoir studies where permeability or porosity are important
properties.
 
One aspect which we did not address in our analysis is the variation
of thermal conductivity with in situ pressure. Sonic velocity varies
strongly with pressure and so does thermal conductivity
\cite{Clauser1995a,Seipold1990}. Attempts have been made to model this
behaviour with respect to the variation of elastic properties with
pressure \cite{Sattel1982,Zimmerman1989}.  Because the Tertiary
samples examined in our work are poorly consolidated, a considerable
pressure effect may be expected. This requires more detailed study in
future.

\section{Acknowledgements}
\label{sec:ackknowledgements}

Our research was funded by the German Federal Environmental Ministry
as part of the activities of its AkEnd Working Group
(http://www.akend.de) through Bundesamt für Strahlenschutz (BFS,
Federal Agency for Radiation Protection), contract no.~9X0009-8390-0
to RWTH Aachen and contract no.~WS~0009-8497-2 to Geophysica
Beratungsgesellschaft mbH. We would like to thank the Geological
Survey of Baden-Württemberg (LGRB, Freiburg) for providing data and
core material. Part of the thermal conductivity measurements were made
at the Leibnitz Institute for Applied Geosciences (GGA, Hannover) by
H.~Deetjen and R.~Schellschmidt.  Sonic velocity measurements under
confining pressure were performed by S.~Mayr (TU Berlin).  F.~Höhne
and D.~Breuer were responsible for the measurements at RWTH
University. The paper was improved by the comments of two anonymous
reviewers. 

\bibliography{ijrm}
\bibliographystyle{elsart-num}

\pb{}
\begin{table}[htbp]
   \centering
  % Table generated by Excel2LaTeX from sheet 'Korrelationen'
\begin{tabular}{|cc|r|r|r|r|}
\hline
                                 \multicolumn{ 6}{|c|}{{\bf $\lambda$(dry)}} \\
\hline
{\bf Property} & {\bf Lithology} & \multicolumn{1}{|c|}{\bf $a_1$} & \multicolumn{1}{|c|}{\bf $a_0$} &  \multicolumn{1}{|c|}{\bf $R$} &  \multicolumn{1}{|c|}{\bf rms} \\
\hline
     $v_p$ & all samples & $ 0.696\pm0.018$ & $-0.485\pm0.050$ &      0.921 &       0.18 \\

  $\rho_b$ & all samples & $ 2.715\pm0.086$ & $-4.167\pm0.183$ &      0.886 &       0.23 \\

    $\phi$ & all samples & $-6.289\pm0.570$ & $ 2.926\pm0.103$ &     -0.849 &       0.24 \\
\hline
     $v_p$ &      sandy & $ 0.744\pm0.028$ & $-0.601\pm0.068$ &      0.898 &       0.17 \\

     $v_p$ &  carbonate & $ 0.680\pm0.031$ & $-0.457\pm0.108$ &      0.893 &       0.20 \\

  $\rho_b$ &      sandy & $ 2.500\pm0.123$ & $-3.740\pm0.250$ &      0.844 &       0.22 \\

  $\rho_b$ &  carbonate & $ 2.942\pm0.179$ & $-4.645\pm0.412$ &      0.851 &       0.24 \\

    $\phi$ &      sandy & $-5.783\pm0.898$ & $ 2.818\pm0.178$ &     -0.772 &       0.25 \\

    $\phi$ &  carbonate & $-6.490\pm0.489$ & $ 2.939\pm0.097$ &     -0.897 &       0.20 \\
\hline
                                 \multicolumn{ 6}{|c|}{{\bf $\lambda$(sat)}} \\
\hline
     $v_p$ & all samples & $ 0.378\pm0.042$ & $ 1.696\pm0.145$ &      0.551 &       0.20 \\

  $\rho_b$ & all samples & $ 2.214\pm0.192$ & $-2.151\pm0.452$ &      0.537 &       0.23 \\

    $\phi$ & all samples & $-3.304\pm0.394$ & $ 3.701\pm0.083$ &     -0.363 &       0.27 \\
\hline
     $v_p$ &      sandy & $ 0.372\pm0.035$ & $ 1.809\pm0.118$ &      0.784 &       0.12 \\

     $v_p$ &  carbonate & $ 0.363\pm0.056$ & $ 1.537\pm0.183$ &      0.449 &       0.16 \\

  $\rho_b$ &      sandy & $ 2.074\pm0.263$ & $-1.713\pm0.618$ &      0.687 &       0.16 \\

  $\rho_b$ &  carbonate & $ 1.696\pm0.230$ & $-1.112\pm0.540$ &      0.548 &       0.16 \\

    $\phi$ &      sandy & $-3.229\pm0.541$ & $ 3.828\pm0.106$ &     -0.696 &       0.16 \\

    $\phi$ &  carbonate & $-2.352\pm0.451$ & $ 3.289\pm0.101$ &     -0.331 &       0.20 \\
\hline
\end{tabular}

%%% Local Variables: 
%%% mode: latex
%%% TeX-master: "ijrm-v03"
%%% End: 

  \caption[Predicting $\lambda$ from different properties.]{Results of
    single regression analysis of thermal conductivity $\lambda$ of
    core samples based on  sonic velocity $v_p$, bulk density
    $\rho_b$, and porosity $\phi$. The top panel gives results for 
    dry, the lower part for  saturated samples. For each  regression
    the fit parameters $a_1$ and $a_0$ (Equation~\ref{eq:linear_fit}),
     with errors, the correlation coefficient $R$, and the rms-error of the fit are
    shown. Fits were computed both for all 
    samples and grouped by lithology. For ease of display units of
    $v_p$ and $\rho_b$ are km\eqdist s\up{-1} and g\eqdist cm\up{-3},
    respectively.} 
  \label{tab:bw_therm_con_pred_results}
 
\end{table}

\pb{}
 
\begin{table}[htbp]
 
    \centering \small{}
% Table generated by Excel2LaTeX from sheet 'Multiple'
\begin{tabular}{|cc|r|r|r|r|r|}
\hline
\multicolumn{1}{|c}{\bf Lithology} & \multicolumn{1}{c|}{\bf
  State} & \multicolumn{1}{|c|}{\bf $a_3$} &
\multicolumn{1}{|c|}{\bf $a_2$} & \multicolumn{1}{|c|}{\bf $a_1$} &
\multicolumn{1}{|c|}{\bf $a_0$} &   \multicolumn{1}{|c|}{\bf rms} \\
\hline
\hline
sandy &        dry & $-0.53\pm0.14$ & $0.615\pm0.096$ &
$0.512\pm0.005$ & $-1.14\pm0.47$ &       0.15 \\ 
sandy &  sat & $0.042\pm0.02$ & $0.504\pm0.053$ &
$0.239\pm0.002$ & $1.07\pm0.27$ &       0.12 \\ 
carbonaceous &        dry & $-2.82\pm0.54$ & $0.209\pm0.031$ &
$0.371\pm0.006$ & $0.64\pm1.06$ &       0.13 \\ 
carbonaceous &  sat & $0.66\pm0.22$ & $1.056\pm0.172$ &
$0.100\pm0.005$ & $-0.11\pm0.83$ &       0.13 \\ \hline\hline
all samples &        dry & $-1.19\pm0.20$ & $0.556\pm0.053$ &
$0.475\pm0.002$ & $-0.73\pm0.26$ &       0.14 \\ 
all samples &  sat & $1.11\pm0.12$ & $0.913\pm0.076$ &
$0.243\pm0.002$ & $-0.22\pm0.38$ &       0.17 \\ 
\hline
\end{tabular}  
\caption{Results of the multiple regression of thermal conductivity
  $\lambda$  versus   sonic velocity $v_p$, bulk density $\rho_b$, and porosity
  $\phi$ using equation~\ref{eq:multi_linear_regression}. For each regression
  coefficients of the fit, coefficient error,  correlation
  coefficient $R$,     and the rms-error of the fit are given. Fits
  were  computed both for all samples (lower part) and grouped by
  lithology (upper part). For ease of display units of 
  $v_p$ and $\rho_b$ are km\eqdist s\up{-1} and g\eqdist cm\up{-3}.}
\label{tab:bw_multi_regression}
 
\end{table}

\pb{}
\begin{table}[htbp]
  \centering
  \begin{tabular}[b]{|ll|ccc|c|}\hline
    \multicolumn{2}{|c|}{\raisebox{-2ex}[0ex][0ex]{Component}} & DT &
    GR & NPHI & $\lambda$ 
    \\\cline{3-6}  
    & & $\mu$s m\up{-1} & API & p.u. & \uwlf{}\\ \hline\hline
    \multicolumn{2}{|c}{\hspace{0ex}Full model} & 
    \multicolumn{4}{c|}{}\\ \hline
    Sand &(quartz) &  182 & 30 & -6 & 7.69 \\
    Shale &(glauconite) & 295 & 150 & 41 & 2.20 \\
    Carbonate &(calcite) & 157 & 11 & 0 & 3.59 \\\hline \hline
    \multicolumn{2}{|c}{\hspace{0ex}Sand/shale model} &
    \multicolumn{4}{c|}{}\\ \hline
    Sand &(quartz) &  N/A & 30 & N/A &  6.39\\
    Shale &(glauconite) & N/A & 150 & N/A & 1.96 \\ \hline
  \end{tabular}
  \caption{Summary of response parameters and component thermal
    conductivities used in the log data analysis. Slowness DT, natural
    gamma radiation GR, and neutron porosity NPHI are used in the
    complex model. For the simple model NPHI is not used and DT
    coefficients are determined from equation~\ref{eq:han1986}.} 
  \label{tab:bw_param_summary_for_tc_predicitive_models}
\end{table}

\pb{}

\pb{}

\pb{}

\pb{}

\pb{}

\pb{}

\pb{}

\pb{}

\pb{}

%\pb{}
%\begin{figure}[htbp]
%  \centering
%  \includegraphics[width=12cm,clip,trim=0 10 0 0]
%  {bw_phi_crossplot.eps}
%  \caption{Crossplot of porosities using the Sand-Shale model with DT
%    and GR as input logs. Circles: Crossplot of DT-GR porosity versus
%    core porosity. Dots: Crossplot of DT-GR porosity versus Neutron
%    porosity.} 
%  \label{fig:bw_phi_crossplot_Han_model}
%\end{figure}

\end{document}